\documentclass[a4paper,fleqn,usenatbib,useAMS]{mnras}

\usepackage{mathptmx}

\usepackage[T1]{fontenc}
\usepackage{ae,aecompl}
\usepackage{graphicx}	
\usepackage{amsmath}	
\usepackage{amssymb}	
\usepackage{url}

\title[Enhanced Fe opacity and hybrid massive pulsators]{The
Impact of Enhanced Iron Opacity on Massive Star Pulsations:
Updated Instability Strips\thanks{To reproduce the results,
all software, opacity tables and the new instability strips are freely available 
for download via 
\url{https://fys.kuleuven.be/ster/Projects/ASAMBA}.
}}

\author[E. Moravveji]{
Ehsan Moravveji$^{1}$\thanks{E-mail: Ehsan.Moravveji@ster.kuleuven.be}
\\
$^{1}$Institute of Astronomy, KU\,Leuven, Celestijnenlaan 200D, 
3001 Leuven, Belgium\\
}

\date{Accepted 2015 September 25. Received 2015 September 6; in original form 2015 August 19}

\pubyear{2015}

\begin{document}
\label{firstpage}
\pagerange{\pageref{firstpage}--\pageref{lastpage}}
\maketitle

\begin{abstract}
Recently, \citet{bailey-2015-01} made a direct measurement of the Iron opacity
at the physical conditions of the solar tachocline.
They found that the wavelength-integrated Iron opacity is roughly 75\% higher
that what the OP and OPAL models predict.
Here, we compute new opacity tables with enhanced Iron and Nickel contributions to the
Rosseland mean opacity by 75\% each, and compute three dense MESA grids of evolutionary models
for Galactic O- and B-type stars covering from 2.5 to 25\,M$_\odot$ from ZAMS until 
$T_{\rm eff}=10\,000$\,K after the core hydrogen exhaustion.
We carry out non-adiabatic mode stability analysis with GYRE, and update the 
extension of 
the instability strips of heat-driven p- and g-mode pulsators, and the hybrid pulsating
SPB\,-\,$\beta$\,Cep stars.
We compare the position of two confirmed late O-type $\beta$\,Cep and eight confirmed 
hybrid B-type pulsators with the new instability domains, and justify that  $\sim$75\% 
enhancement, only in Iron opacity, is sufficient to consistently reproduce the observed
position of these stars on the $\log T_{\rm eff}$ versus $\log g$ plane.
We propose that this improvement in opacities be incorporated in the input physics 
of new stellar models.
\end{abstract}


\section{Introduction}\label{s-intro}
The stellar opacity is a key ingredient of our simplified 1D stellar structure, 
evolution and pulsations models; 
it determines the efficiency of
radiative energy transfer, and controls the luminosity, hence lifespan of stars.
Indeed, any improvement in our understanding of stellar opacities impacts the morphology of 
evolutionary tracks, the distribution of basic thermodynamical quantities 
(such as temperature, pressure and density) inside the models, and changes the 
theoretical pulsation frequencies.
It also shifts the theoretical isochrones, and consequently the ages of stellar populations.

Recently, \citet{bailey-2015-01} conducted a direct measurement of the Iron opacity at 
the physical conditions resembling the solar tachocline ($T=$1.91 -- 2.26$\times10^6$ K, 
$n_e=$0.7 -- 4$\times10^{22}$ cm$^{-3}$).
Their comparison between the wavelength-dependent Fe opacity and those computed from the 
Opacity Project \citep[OP,][]{seaton-1994-01,badnell-2005-01} 
showed that the models 
underestimate Fe opacity by 30\% to 400\% (their Fig.\,3a and 3c), leading to the fact 
that the measured Iron Rosseland mean opacity (see below) is $\beta_{\rm Fe}\approx1.75$
times larger than the OP models.
Moreover, Nickel is the 2$^{\rm nd}$ abundant Iron-group element in typical 
main sequence (MS) stars.
The similarity of the atomic structures of Fe and Ni has already allowed OP and OPAL teams
to construct Ni monochromatic opacities by scaling Fe opacities 
\citep{badnell-2005-01,iglesias-2015-01}.
This implies that the Ni monochromatic opacity could be underestimated by roughly the same
factor $\approx1.75$ as measured for Fe.
Thus, there is still enough room for improving Ni opacity.
It is noteworthy that the recent revised OPAL \citep{iglesias-2015-01} opacities for Fe
reasonably agrees with the OP ones and is still significantly below the measurements.
The improvements provided by \citeauthor{bailey-2015-01} prompts an update to the opacity tables before any further stellar evolutionary models are computed.

One of the direct implications of Fe opacity enhancement is its immediate influence on
destabilizing low-order p- and high-order g-modes in $\beta$\,Cep and Slowly Pulsating B
(SPB) stars, where
the Iron-opacity peak at $\log T\simeq5.2$ K is already known to be responsible for
\citep{gautschy-1993-01,dziembowski-1993-01,dziembowski-1993-02}.
The height of this opacity peak depends directly on 
(a) the assumed mixture, i.e. the relative abundance of the Iron-group elements, 
Fe, Ni, Co, Cr, and Mn \citep{salmon-2012-01}, 
(b) the assumed metallicity (e.g. $Z=0.014$ versus $Z=0.020$), and
(c) the contribution of each element to the net opacity $\kappa_{\rm net}(\nu)$.
Conseqeuently, with the increasing role of Iron-group elements in the Rosseland mean opacity
(Sect.\,\ref{s-rosseland}), the number of excited p- and g-modes increases, in additoin
to their instability domains on the Kiel ($\log T_{\rm eff}$ versus $\log g$) diagram.
Most importantly, the overlapping region between the two instability strips 
-- where the so-called hybrid pulsators lie -- depends critically on the computations of
Fe and Ni radiative opacities.
In this paper, we prepare new opacity tables with different Fe and Ni contributions 
(Sect.\,\ref{s-rosseland}), and employ them to compute three evolutionary grids
(Sect.\,\ref{s-inputs}) and their corresponding instability strips of the upper HR diagram
(Sect.\,\ref{s-strip}).
By incorporating the Iron opacity enhancement, we solve the discrepancy between the predicted
instability domain of hybrid SPB\,-\,$\beta$\,Cephei pulsators and those of detected hybrid 
B pulsators (Sect.\,\ref{s-obs-vs-IS}).

\section{Enhanced Monochromatic Iron Opacity and the Rosseland Mean}\label{s-rosseland}
The radiative opacity contains contributions from electron scattering $\kappa_{\rm es}$, 
bound-bound inter-shell 
$\Delta n\geq0$ 
transitions $\kappa_{\rm bb}$, bound-free transitions 
including photo-ionizations $\kappa_{\rm bf}$, and free-free transitions $\kappa_{\rm ff}$.
For a given photon frequency $\nu$, the combined monochromatic photon absorption cross section 
for any element $i$ is designated by $\kappa^i(\nu)$, and is summed as follows 
\begin{equation}\label{e-kappa-nu}
\kappa^i(\nu) = 
  \beta_i \,\left(\kappa^i_{\rm bf}+\kappa^i_{\rm ff}+\kappa^i_{\rm bb}\right)
  \left[1-e^{-h\nu/K_{\rm B}T  }\right],
\end{equation}
where $K_{\rm B}$ is the Boltzman constant.
The net frequency-dependent opacity $\kappa_{\rm net}(\nu)$ is a sum over $N$ available 
elements, i.e. 
$\kappa_{\rm net}(\nu)=\kappa_{\rm es}+\sum_{i=1}^{N}\kappa^i(\nu)$.
The Rosseland mean opacity, here $\kappa$, is a harmonic mean of the reciprocal 
opacity of stellar material weighted by the Planck energy distribution function 
$B(\nu, T)$ and integrated over frequency (or wavelength)
\begin{equation}\label{e-ross}
\frac{1}{\kappa} = \frac{1}{dB/dT}\int_0^\infty 
       \frac{1}{\kappa_{\rm net}(\nu)}\frac{dB(\nu, T)}{dT}\,d\nu.
\end{equation}
Note that $dB/dT=\int_0^\infty d\nu\,dB(\nu, T)/dT=acT^3/\pi$ where $a$ and $c$ are the 
radiation constant and speed of light, respectivley.

We introduced an additional vector of enhancement factors $\beta_i$ 
in Eq.\,(\ref{e-kappa-nu}) that allows modifying the contributions of individual 
elements to the net opacity before carrying out the Rosseland mean in Eq.\,\ref{e-ross}.
Setting all factors to unity, $\beta_i$=1, reproduces the default monochromatic
OP opacities.
We choose to set $\beta_{\rm Fe}=1.75$ as suggested by \citet{bailey-2015-01}, and compare
the resulting Rosseland mean with the default case where $\beta_{\rm Fe}=1.0$.
Additionally, we predict mode instability properties for the case $\beta_{\rm Ni}=1.75$.

The computation of monochromatic (Eq.\,\ref{e-kappa-nu}) and the Rosseland mean  
(Eq.\,\ref{e-ross}) opacities are possible by using the public 
\href{http://sourceforge.net/projects/mesa/files}{\tt OPCD\,3.3}
\citep{seaton-2005-01,badnell-2005-01}.
\citet{hu-2010-01} have already implemented an interface to call 
\href{http://sourceforge.net/projects/mesa/files}{\tt OPCD}
from the 
MESA stellar structure and evolution code \citep{paxton-2011-01, paxton-2013-01, 
paxton-2015-01} that we also use.
Currently, \href{http://sourceforge.net/projects/mesa/files}{\tt OPCD}
uses atomic data for 17 elements (hence $i$=H, He, C, N, O, Ne, Na, Mg,
Al, Si, S, Ar, Ca, Cr, Mn, Fe and Ni).
We developed additional routines that call the \texttt{chem}, \texttt{eos} 
and \texttt{kap} modules in MESA to compute $\kappa$ for any assumed $\beta_i$ 
over a pre-defined range of temperatures and densities.
The output is a set of ASCII tables identical to the OPAL Type\,{\tt I} tables 
which are commonly used for interpolation in 1D stellar evolution codes.

\begin{figure}
\includegraphics[width=\columnwidth]{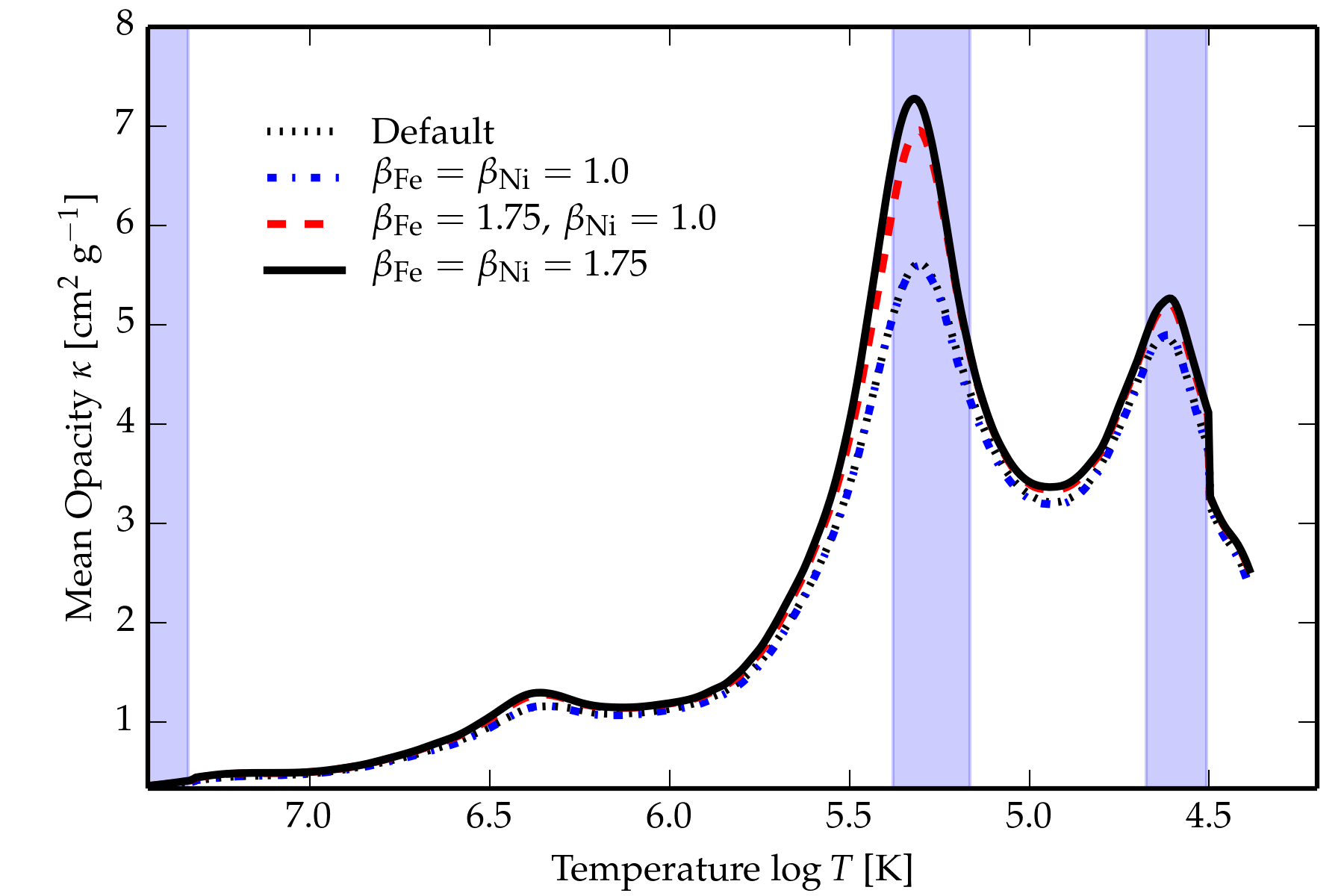}
\caption{Rosseland opacity $\kappa$ in a 10\,M$_\odot$ star at fixed center hydrogen
mass fraction $X_{\rm c}=0.60$ and $Z=0.014$.
The profiles for the default MESA tables (black dotted) and 
$\beta_{\rm Fe}=\beta_{\rm Ni}=1.0$ (blue dash-dot) are identical.
The red dashed and black solid curves correspond to 
($\beta_{\rm Fe},\,\beta_{\rm Ni}$)\,$=$\,(1.75, 1.0), and (1.75, 1.75), respectively.
The convective regions are highlighted in blue.}
\label{f-kappa}
\end{figure}
%

\section{Stellar models}\label{s-inputs}
We computed new sets of opacity tables by adopting the \citet{asplund-2009-01} solar mixture 
for three combinations of ($\beta_{\rm Fe},\,\beta_{\rm Ni}$)\,$=$\,(1.0, 1.0),
(1.75, 1.0) and (1.75, 1.75).
Fig.\,\ref{f-kappa} compares the resulting $\kappa$ profiles using the 
default MESA OP opacity tables, and the new tables with 
($\beta_{\rm Fe},\,\beta_{\rm Ni}$)\,$=$\,(1.0, 1.0), (1.75, 1.0) and (1.75, 1.75), 
respectively.
The first two $\kappa$ profiles are identical, demonstrating that our new 
$\beta_{\rm Fe}=1.0$ tables are consistent with the default MESA tables.
By increasing Fe and Ni opacity factors $(\beta_{\rm Fe},\,\beta_{\rm Ni})$ 
to (1.75,\,1.0) and (1.75,\,1.75), 
the height of the Iron-bump increases by $\sim24\%$ and $\sim30\%$, respectively.
Similarly, the total photon interaction cross section, i.e. $\sigma^2=\int_0^{M_\star}\kappa\,dm$ [cm$^2$] increases by 5.11\% and 5.57\%, respectively,
when compared to the case of $(\beta_{\rm Fe},\,\beta_{\rm Ni})=(1.0,\,1.0)$. 
Here, $M_\star$ and $dm$ are the total mass and mass increments.

\begin{figure*}
\begin{minipage}{0.48\textwidth}
\includegraphics[width=\columnwidth]{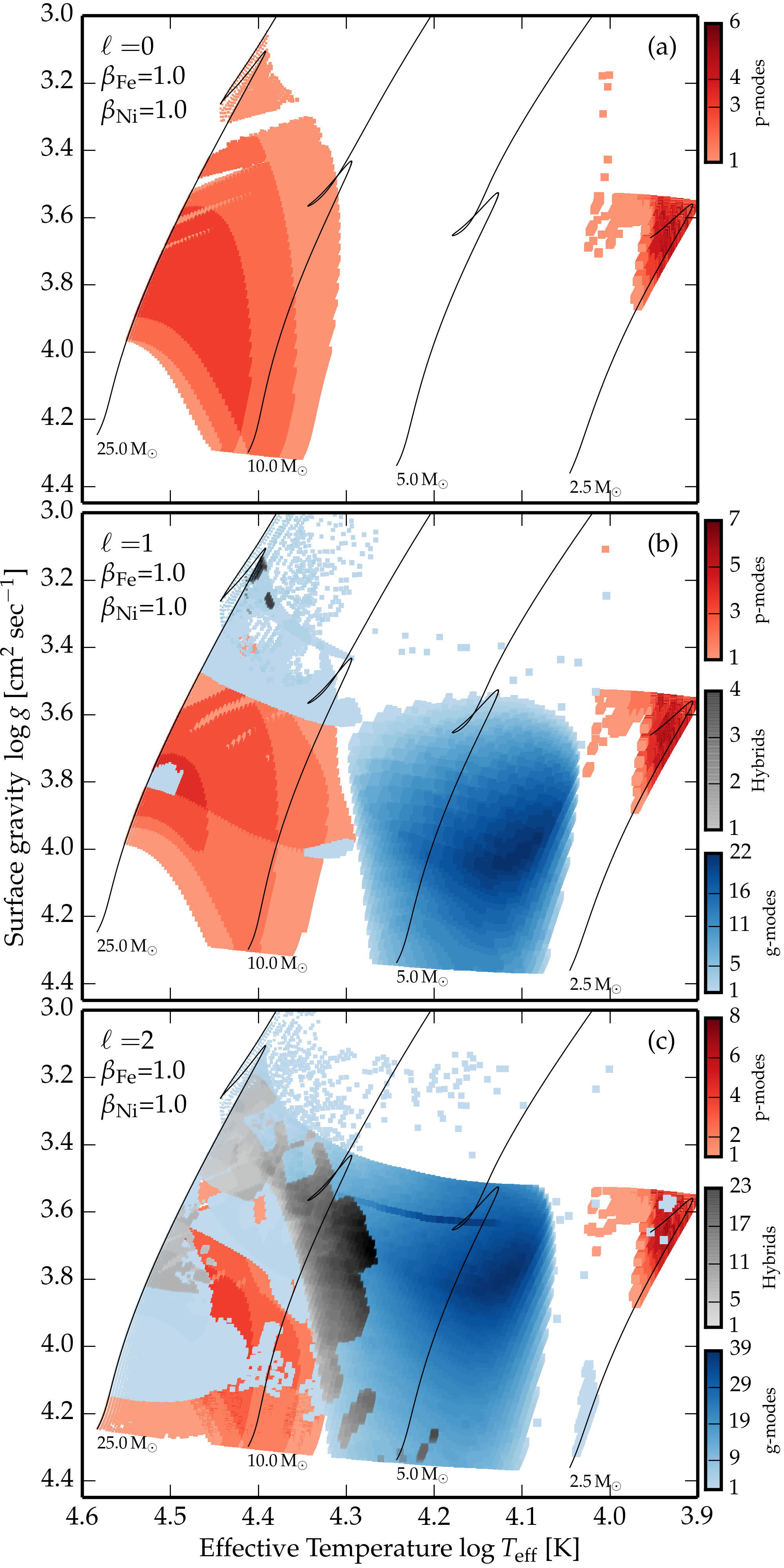}
\end{minipage}
\hspace{0.01\textwidth}
\begin{minipage}{0.48\textwidth}
\includegraphics[width=\columnwidth]{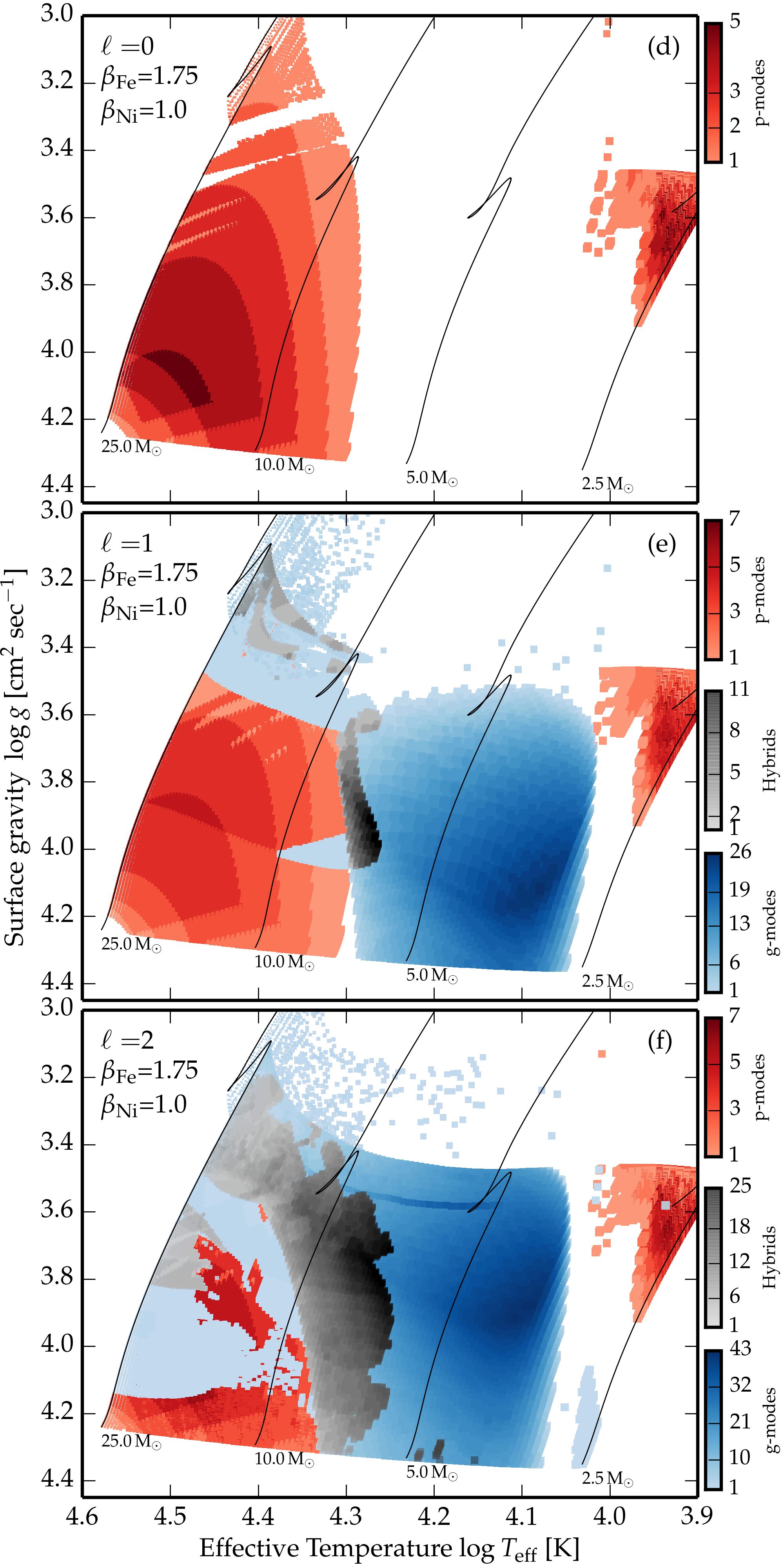}
\end{minipage}
\caption{Instability strips of pure p-modes (red), pure g-modes (blue), and hybrid 
        (grey) pulsators.
        Top, middle, and bottom panels show radial, dipole, and quadrupole modes, 
        respectively.
        The left panels (a, b, c) correspond to tracks computed with 
        $\beta_{\rm Fe}=\beta_{\rm Ni}=1.0$, and similarly the right panels 
        (d, e, f) correspond to $\beta_{\rm Fe}=1.75$ and $\beta_{\rm Ni}=1.0$.
        All evolutionary tracks are computed with fixed metallicity $Z=0.014$,
        the \citet{asplund-2009-01} mixture, and exponential core overshooting 
        $f_{\rm ov}=0.02$.
        Four tracks are shown for comparison.
        Tracks are computed only between 2.5 to 25\,M$_\odot$.
        Colors encode the number of unstable modes.
        Refer to the color figure for better visibility.}
\label{f-IS}
\end{figure*}

Then, we plugged the new opacity tables into MESA to compute three evolutionary grids 
of 101 evolutionary tracks each, uniformly spaced between 2.5 and 25\,M$_\odot$ in 
logarithmic scale.
We adopted the Galactic B-star composition of \citet{nieva-2012-01} 
$(X_{\rm ini}, Z_{\rm ini})=(0.71, 0.014)$, and the \citet{asplund-2009-01} mixture.
We include mass loss to remain consistent with observation, although it has a stabilizing
effect on heat-driven modes \citep{godart-2009-01}.
An exponential core overshoot of $f_{\rm ov}=0.02$ is also added.
The evolution started from zero-age-main-sequence (ZAMS), and is terminated once 
the models reach $T_{\rm eff}=10\,000$\,K after core hydrogen depletion.
We stored equilibrium models for every $\sim0.001$ drop in center 
hydrogen $X_{\rm c}$ during the MS phase, and by 100\,K change in $T_{\rm eff}$ during
post-MS phase.
For every model, we used GYRE \citep[][v.4.2]{townsend-2013-01} to compute non-adiabatic
frequencies of radial ($\ell=0$), dipole ($\ell=1$) and quadrupole ($\ell=2$) zonal 
modes ($m=0$) in the frequency range 0.4 to 20\,d$^{-1}$.
We discriminate between p-modes ($n_{\rm pg}\geq0$) and
g-modes ($n_{\rm pg}<0$) based on the net radial order $n_{\rm pg}=n_{\rm p}-n_{\rm g}$;
here, $n_{\rm p}$ (or $n_{\rm g}$) is the number of p- (or g-) dominated radial nodes.
The excited modes are distinguished by considering the sign of the imaginary part
of the eigenfrequencies \citep{unno-1989-book}, and then enumerated.
\citet{dziembowski-2008-01} define hybrid pulsators as those which exhibit $\beta$\,Cep
variability simultneously with SPB type.
However, there is no clear-cut definition of low-order versus high-order modes in terms 
of $n_{\rm pg}$, and one must set a definition.
We identify a model as a hybrid, once it has at least one unstable low-order p- and/or 
g-mode ($-2\leq n_{\rm pg}\leq+2$), in addition to at least one unstable high-order
g-mode ($n_{\rm pg}<-3$).
This allows classifying each model into being a pure p-mode, pure g-mode, or 
hybrid pulsator.

\section{Revised Instability Strips of the Upper HR Diagram}\label{s-strip}

Fig.\,\ref{f-IS} shows the instability strips of the unstable radial (top), 
dipole (middle), and quadrupole (bottom) modes using $\beta_{\rm Fe}=\beta_{\rm Ni}=1.0$
tables (left) and $\beta_{\rm Fe}=1.75, \,\beta_{\rm Ni}=1.0$ tables (right).
The instability strip for $\beta_{\rm Fe}=\beta_{\rm Ni}=1.75$ grid is quite similar 
to that of $\beta_{\rm Fe}=1.75$, and not shown here.

The immediate consequence of increasing $\beta_{\rm Fe}$ from 1.0 
to 1.75 is an increase in the number of unstable g-modes and hybrids.
In Fig.\,\ref{f-IS}a there is a gap at higher mass near the ZAMS for 
$\log T_{\rm eff}\gtrsim4.45$, where no unstable modes are predicted 
\citep{pamyatnykh-1999-01,saio-2011-01,walczak-2015-01}.
However, this region is filled up in Fig.\,\ref{f-IS}d by few low-order radial modes
thanks to $\beta_{\rm Fe}=1.75$.
The same is true for dipole p-modes.
Thus, all stars more massive than $\sim7\,$M$_\odot$ exhibit $\beta$\,Cep type variability
as soon as they reach the ZAMS.
In Fig.\,\ref{f-IS}b, the SPB instability strip is separated from the 
$\beta$\,Cep strip, while in Fig.\,\ref{f-IS}e the two regions smoothly merge, 
giving rise to a much extended hybrid instability domain.
This is because the $\beta$\,Cep instability strip has now stretched to lower 
$T_{\rm eff}$, and the SPB strip has extended to higher and lower $T_{\rm eff}$.
The quadrupole g-mode instability domain in Figs.\,\ref{f-IS-obs}c and \ref{f-IS-obs}f 
seems to be a common pulsation feature of all
massive O- and B-type stars, and the role of Fe contribution to the opacity has a 
sizeable influence on the widening of the hybrid region towards lower effective 
temperatures and higher surface gravities.

The SPB instability strip hosts rich high-order dipole and quadrupole g-mode pulsators.
The success of asteroseismic modelling of SPB stars \citep[e.g.][]{moravveji-2015-01} 
depends a\,priori on identifying consecutive series of g-modes with the same degree 
$\ell$ that form a period spacing, with a possible deviation from the asymptotic spacing.
Such modes probe the extent, thermal and chemical structures of the overshooting layer 
on top of the convective core, in addition to the $\mu$-gradient layer
\citep{moravveji-2015-02}.
Thus, SPBs are promising asteroseismic targets for space observations, 
and should be carefully selected.
The dark blue patches in Figs.\,\ref{f-IS}b, \ref{f-IS}c, \ref{f-IS}e and \ref{f-IS}f
mark the position of most unstable g-modes with initial mass $\sim3.5$\,M$_\odot$.
This clearly explains the recent discovery of the two \textit{Kepler} SPB stars, namely
KIC\,10526294 \citep[$\sim$3.25\,M$_\odot$,][]{papics-2014-01,moravveji-2015-01} 
and KIC\,7760680 \citep[$\sim$3.3\,M$_\odot$,][]{papics-2015-01}, with 19 and 36 
identified consecutive dipole g-modes, respectively.
At higher surface gravities, such stars are first rich dipole pulsators, which later on,
turn into rich quadrupole pulsators (Figs.\,\ref{f-IS}e and \ref{f-IS}f).
Moreover, this richness explains the large number of observed significant 
low-frequency peaks in the g-mode frequency range in numerous additional 
{\it Kepler\/} SPB candidates yet to be analysed in full detail.
The predicted position of the richest SPB stars (dark blue patches in Fig.\ref{f-IS}) 
is a roadmap for selecting future targets for the on-going BRITE-constellation, the
two-wheel \textit{Kepler} (abbreviated K2), and the future PLATO space missions.

Below the SPB instability strip where the 
partial He ionization region destabilizes the high-order $\delta$\,Scuti-type
variability \citep{pamyatnykh-1999-01}, we find unstable low-degree p-modes and 
quadrupole g-modes.
Even though this is below our region of attention, but it is worth to mention that 
recently \cite{balona-2015-01} predicted that a factor 2 increase in the Fe opacity around 
$\log T\sim5.06$ can destabilize g-modes and allows explaining the observed 
low-frequency peaks in their \textit{Kepler} sample of A- and F-type $\delta$\,Scuti
pulsators.
Thus, our results agrees with their predictions, too.

\section{Observed and predicted domains of hybrid pulsators}\label{s-obs-vs-IS}
To manifest the consistency of the enhanced Fe opacity with the observed instability
domains of heat-driven massive pulsators, we make two comparisons between the 
observed position of two confirmed 
late O-type $\beta$\,Cep pulsators, and all confirmed SPB\,-\,$\beta$\,Cep hybrids 
from the literature.
Table\,\ref{t-stars} gives a summary of these carefully selected targets.

\begin{table}
	\centering
	\caption{List of confirmed SPB\,-\,$\beta$\,Cep pulsators.
	The given ID is the numbers used to mark each star in Fig.\,\ref{f-IS-obs}.
    }
	\label{t-stars}
	\begin{tabular}{lllll}
	\hline
		Star & ID & $\log T_{\rm eff}$ & $\log g$            & Var.  \\
		Name &    & [K]                & [cm/sec$^2$]        & Type  \\
		\hline
		HD\,46202$^{\rm(a)}$       & 1 & 4.533$\pm$0.005 & 4.18$\pm$0.05 & $\beta$\,Cep \\
		EPIC\,202060092$^{\rm(b)}$ & 2 & 4.544$\pm$0.050 & 4.50$\pm$0.50 & $\beta$\,Cep \\ 
		\hline
		$\gamma$\,Peg$^{\rm(a)}$   & 1 & 4.325$\pm$0.026 & 4.15$\pm$0.15 & Hybrid  \\ 
		$\nu$\,Eri$^{\rm(a)}$      & 2 & 4.371$\pm$0.018 & 3.75$\pm$0.15 & Hybrid  \\
		12\,Lac$^{\rm(a)}$         & 3 & 4.389$\pm$0.017 & 3.65$\pm$0.15 & Hybrid  \\ 
		16\,Mon$^{\rm(c)}$         & 4 & 4.301$\pm$0.021 & 4.20$\pm$0.10 & Hybrid  \\
		V1449\,Aqu$^{\rm(a)}$      & 5 & 4.389$\pm$0.026 & 3.83$\pm$0.30 & Hybrid  \\ 
		HD\,50230$^{\rm(a)}$       & 6 & 4.255$\pm$0.035 & 3.80$\pm$0.30 & Hybrid  \\
		HD\,43317$^{\rm(d)}$       & 7 & 4.225$\pm$0.025 & 3.90$\pm$0.10 & Hybrid  \\
		HD\,170580$^{\rm(e)}$      & 8 & 4.301$\pm$0.021 & 4.10$\pm$0.15 & Hybrid  \\
		\hline
	\end{tabular}
	\begin{flushleft}
	(a) \citet{aerts-2013-01} and references therein,
	(b) \citet{buysschaert-2015-01}: the star has a poorly known fundamental 
	    parameters due to its binary nature, poor quality spectra and inadequate 
	    orbital monitoring; thus, we assumed 2\,000\,K uncertainty in $T_{\rm eff}$ 
	    and 0.5\,dex in $\log g$,
	(c) \citet{thoul-2013-01},
	(d) \citet{briquet-2013-01}: it is the only magnetic and rapid rotator 
	    (50\% critical) here, and its position on the Kiel diagram is less secure,
	(e) CoRoT target (Conny Aerts, private communication).
	\end{flushleft}
\end{table}

\begin{figure*}
\begin{minipage}{0.48\textwidth}
\includegraphics[width=\columnwidth]{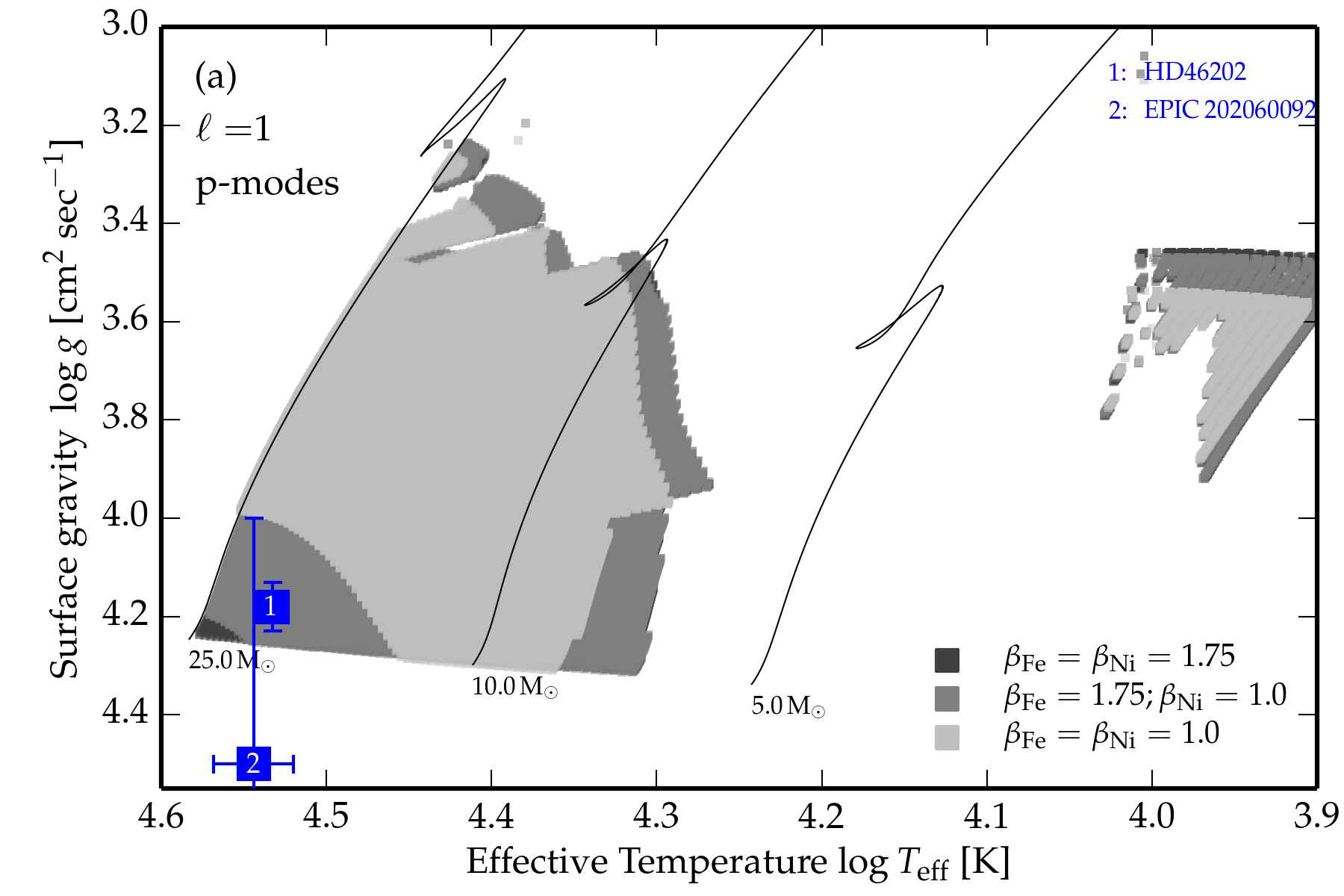}
\end{minipage}
\hspace{0.01\textwidth}
\begin{minipage}{0.48\textwidth}
\includegraphics[width=\columnwidth]{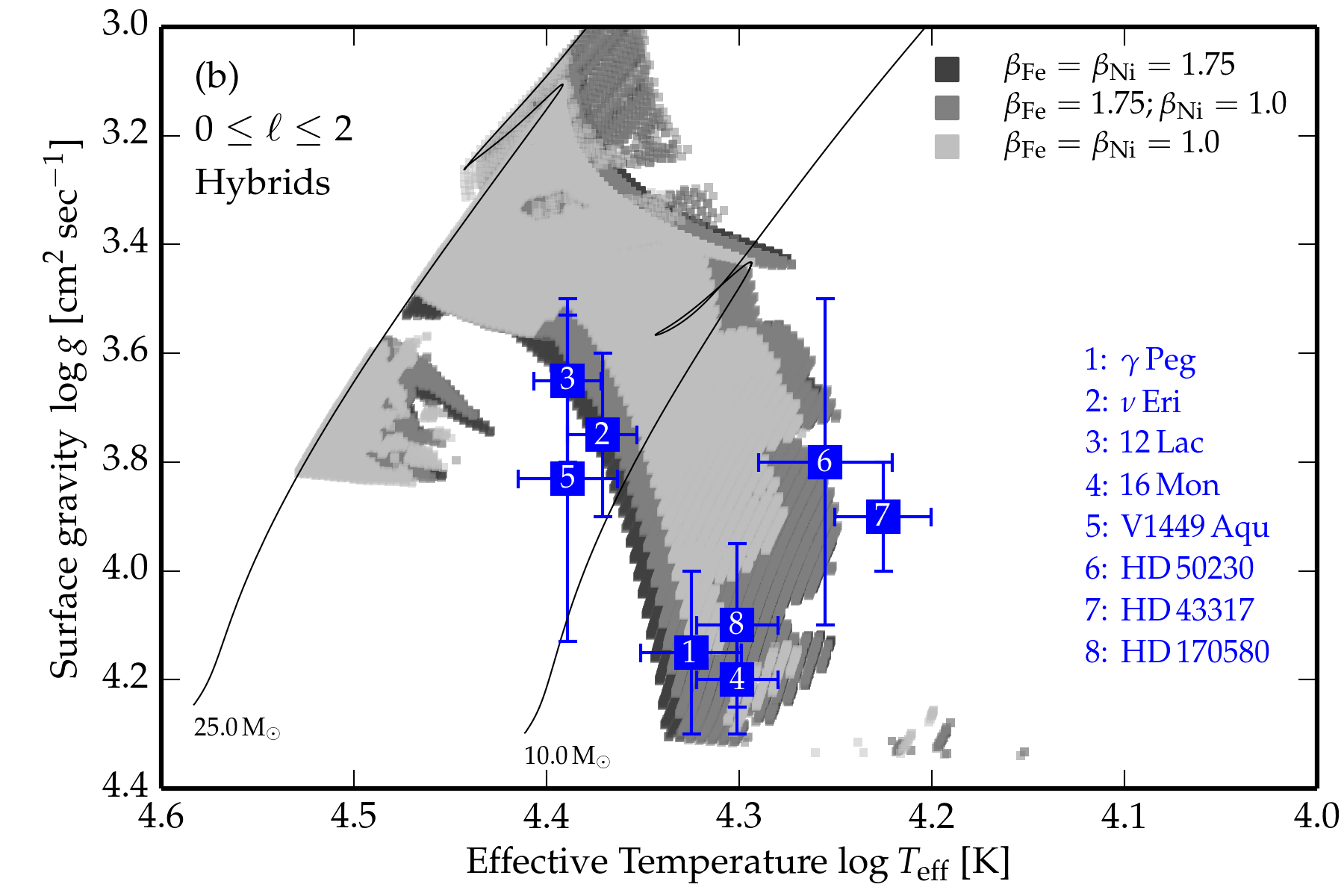}
\end{minipage}
\caption{The extension of dipole p-mode (left) and low-degree hybrid instability strips
for massive OB-type stars on the Kiel diagram.
The light grey corresponds to ($\beta_{\rm Fe},\,\beta_{\rm Ni}$)=(1.0,\,1.0), the 
medium grey corresponds to (1.75, 1.0) and the dark grey corresponds to (1.75, 1.75).
Clearly, Ni opacity enhancement has minute impact on extending the instability domains 
in both cases.
The blue squares show the position of confirmed $\beta$\,Cep (left) and hybrid
pulsators (right).
Table\,\ref{t-stars} gives an overview of these targets.}
\label{f-IS-obs}
\end{figure*}

Fig.\,\ref{f-IS-obs} shows the instability region of dipole p-modes (left), and 
hybrids (right) from the three 
($\beta_{\rm Fe},\,\beta_{\rm Ni}$) combinations of (1.0,\,1.0) in light grey, 
(1.75,\,1.0) in medium grey, and (1.75,\,1.75) in dark grey.
In Fig.\,\ref{f-IS-obs}a, an increase in $\beta_{\rm Fe}$ (and to lesser extent 
$\beta_{\rm Ni}$) extends the instability domain of pure dipole p-modes, and fills up the
gap at the high temperature and surface gravity region up to the ZAMS.
This explains the presence of the only two observed $\beta$\,Cep pulsators in this gap, 
which were formerly unexplained.
Unfortunately, there exists no more confirmed late O-type $\beta$\,Cep stars in 
this extreme region, and future K2, BRITE-constellation and PLATO observations can 
hopefully improve the picture.
In Fig.\,\ref{f-IS-obs}b, the Fe opacity enhancement unambiguously extends the 
hybrid instability domain by $\sim0.2$\,dex to higher surface gravities, and by 
$\sim0.1$\,dex to lower $T_{\rm eff}$.
Thus, the eight SPB\,-\,$\beta$\,Cep hybrids either fall inside the hybrid domain, 
or are much closer to the boundary compared to the case $\beta_{\rm Fe}=1.0$.
As an example, Fig.\,\ref{f-growth-rates} demonstrates how unstable 
modes can be destabilized by increasing $\beta_{\rm Fe}$ and $\beta_{\rm Ni}$ for 
the first three hybrids in Table\,\ref{t-stars}.

\section{Conclusions}\label{s-conclude}
The long-standing difficulty to destabilize heat-driven modes that were present in the 
observations but not predicted in theory can be accurately resolved by enhancing the 
Iron opacity by $\sim$75\% in the models.
This factor comes from direct opacity measurement of \citeauthor{bailey-2015-01}, 
and better explains the distribution of ten $\beta$\,Cep and hybrid pulsators on the 
Kiel diagram.
The required increase in the Z-bump was correctly predicted by many authors
such as \cite{dziembowski-2008-01} for $\nu$\,Eri (Table\,\ref{t-stars}) 
and \citet{salmon-2012-01} for the LMC variables.
Thus, {\sl the input physics of stellar models should now adapt to $\beta_{\rm Fe}=1.75$ }
and consider more opaque stellar plasma, to provide a more realistic picture of 
massive stellar structure, evolution, and pulsation. 

During the evolution of SPBs, the number of excited modes evolves as well. 
This flags late B-type stars of $\sim$\,3M$_\odot$ to 4\,M$_\odot$ as the most 
promising asteroseisimic
targets for the on-going space based photometry by the K2, and BRITE-constellation, in
addition to the future PLATO missions.
Such observations will deliver input to distinguish between different types of 
observed variability in OB-type stars, such as rotational modulation, 
excitation by $\kappa$-mechanism, stochastic excitation of p-modes
\citep{degroote-2010-02}, and excitation by internal gravity waves \citep{aerts-2015-02}.

\section*{Acknowledgements}
{\footnotesize 
E.~Moravveji thanks Conny Aerts (KU Leuven) and  Cole Johnston (KU Leuven) for reading the manuscript, 
and is grateful to Haili Hu (SRON), Aaron Dotter (ANU), 
Peter P\'{a}pics (KU Leuven), Bill Paxton (UCSB), and Geert Jan Bex (UHasselt) 
for valuable discussions and support.
The research leading to these results has received funding from the People Programme (Marie
Curie Actions) of the European Union's Seventh Framework Programme FP7/2007-2013/ under REA
grant agreement n$^\circ$\,623303 for the project ASAMBA.
The computational resources and services used in this work were provided by the 
VSC (Flemish Supercomputer Center), funded by the Hercules Foundation and the Flemish
Government - department EWI.}

\bibliographystyle{mnras}
\bibliography{my-bib.bib} 
\appendix

\section{Growth Rates}

\begin{figure}
\includegraphics[width=\columnwidth]{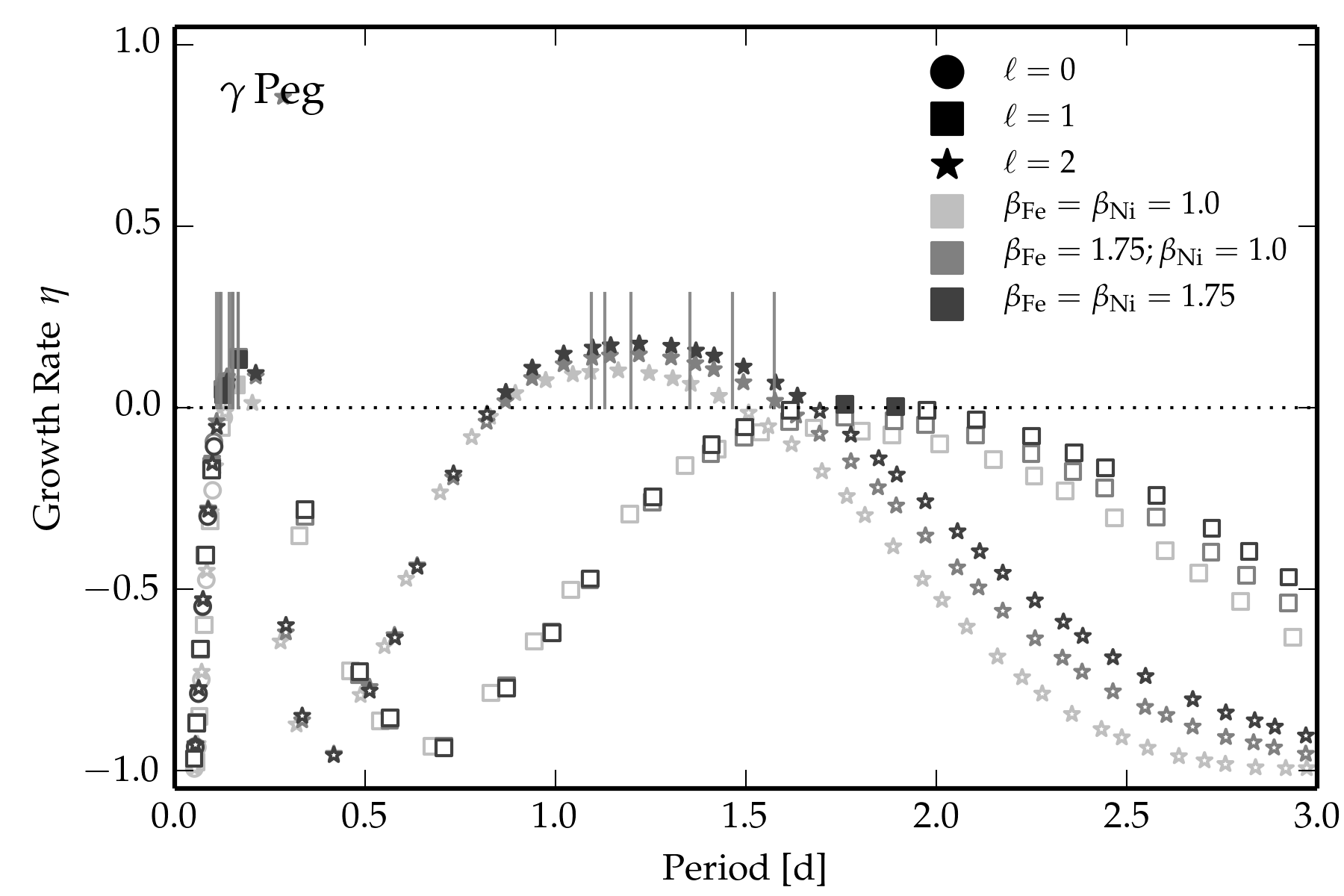}
\includegraphics[width=\columnwidth]{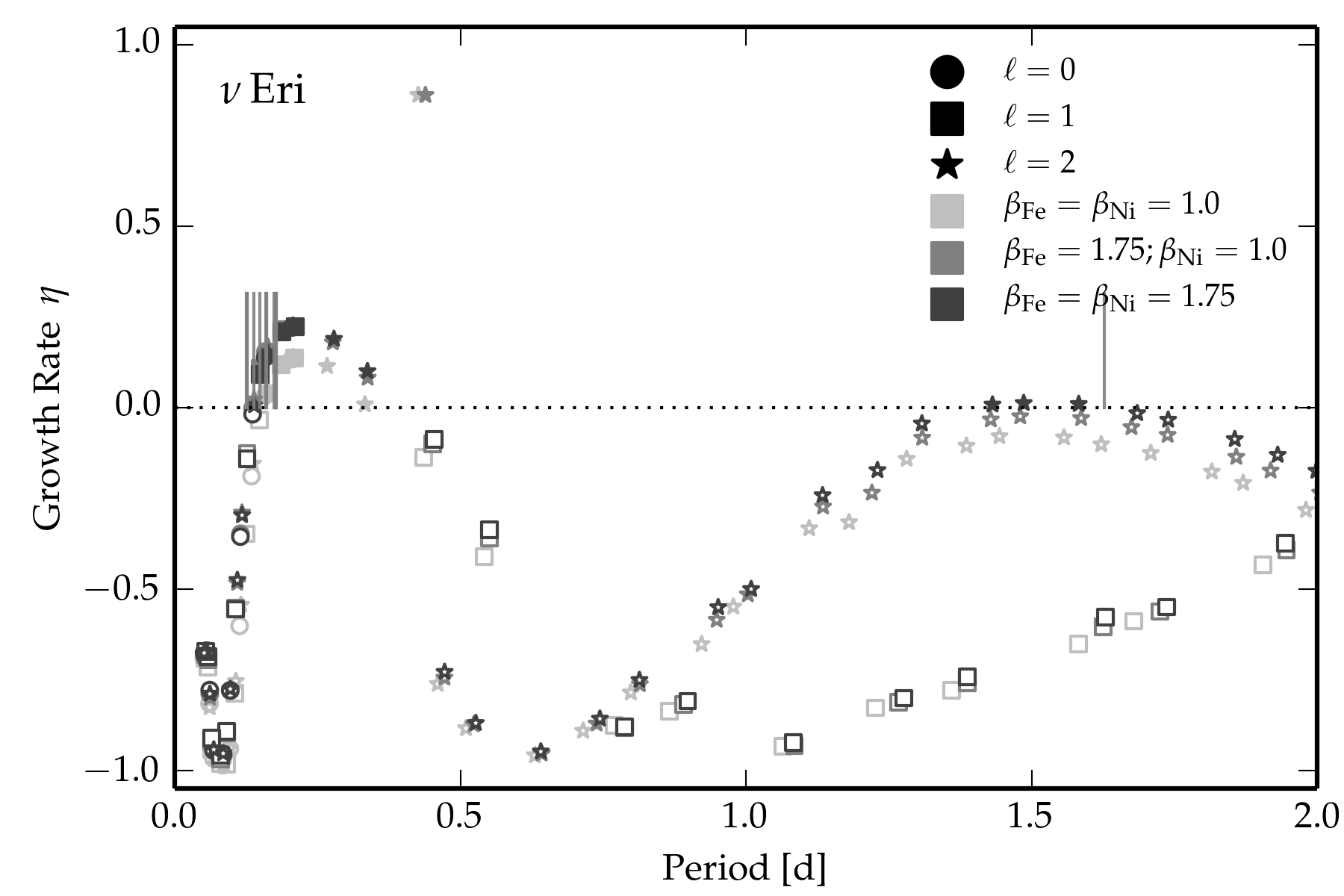}
\includegraphics[width=\columnwidth]{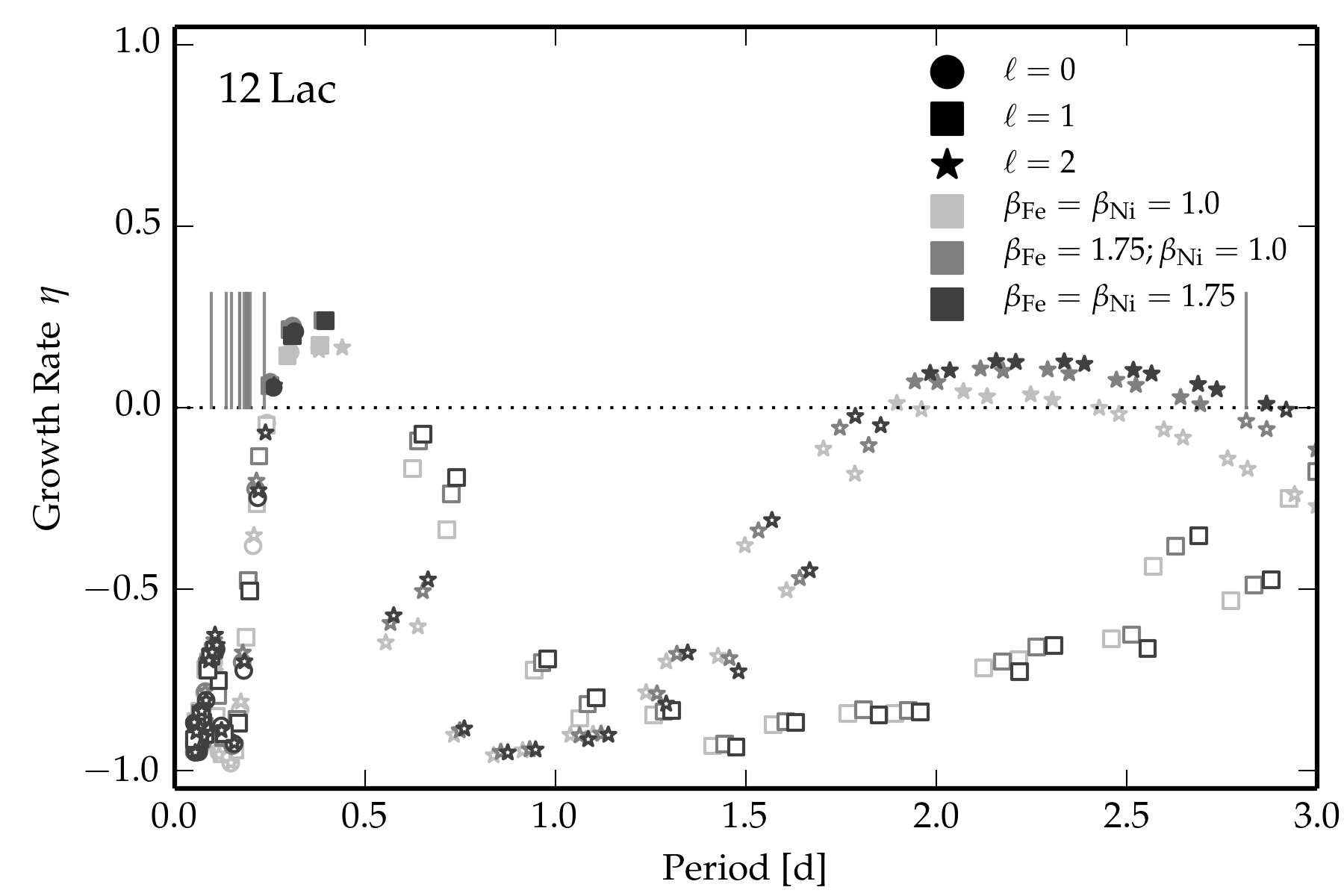}
\caption{Theoretical growth rate $\eta$ for $\gamma$\,Peg (top), $\nu$\,Eri (middle)
and 12\,Lac (bottom).
For each object, the input models are close to their spectroscopic $T_{\rm eff}$ and 
$\log\,g$ (see Table\,\ref{t-stars}).
The vertical lines mark the observed period of each star.
}
\label{f-growth-rates}
\end{figure}

The normalized growth rate $\eta=W/\int_0^{R_\star}|dW/dr|\,dr$ predicts if a mode is 
stable ($W\leq0$) or unstable ($W>0$) against destabilizing effect of the 
$\kappa$-mechanism.
$W$ and $dW/dr$ are the net and differential work, respectively \citep{unno-1989-book}.
Increasing $\beta_{\rm Fe}$ and $\beta_{\rm Ni}$ is expected to destabilize the marginally
stable modes, give rise to larger number of excited modes, and consequently 
extend the instability domains.
We take the first three stars from Table\,\ref{t-stars}, and select three representative
models for each star from our grids close to their position on the Kiel diagram.
The initial mass of these models agree with the literature.
The growth rates are shown in Fig.\,\ref{f-growth-rates}.
For $\nu$\,Eri and 12\,Lac, the more opaque models can better explain their high-order 
g-mode.
Note that the exact reproduction of the observed frequencies of each star is 
beyond the purpose of this \textit{Letter};
we only intend to show that the enhanced Fe and Ni opacities can better explain 
the number of observed modes in these hybrid pulsators, specifically their high-order 
g-modes.


\bsp	
\label{lastpage}
\end{document}